\documentclass[prb,reprint, superscriptaddress, nofootinbib, amsmath,amssymb, aps,longbibliography]{revtex4-1}

\usepackage{graphicx}
\usepackage{dcolumn}

\usepackage{bm}
\usepackage{caption}
\usepackage{float}
\usepackage{lipsum}
\usepackage{amsmath}
\usepackage{amsfonts}
\usepackage{mathrsfs}
\usepackage{mathtools}
\usepackage{amssymb}
\usepackage{dsfont}
\usepackage{color}
\usepackage{tikz}
\usetikzlibrary{automata, positioning, arrows}
\usepackage{nccmath}
\usepackage{caption}
\usepackage{subcaption}
\usepackage[utf8]{inputenc}
\usepackage{babel}
\usepackage{amsmath,amsthm,amssymb}
\usepackage{graphicx}
\usepackage{physics}
\usepackage{listings}
\usepackage{enumitem}
\usepackage{microtype}
\usepackage[raiselinks=true, linktoc=all,hidelinks]{hyperref}
\usepackage[capitalise]{cleveref}
\usepackage{placeins}
\newcolumntype{L}{>{$}l<{$}}

\allowdisplaybreaks[1]
\newcommand{\av}[1]{\left\langle #1 \right\rangle}
\newcommand{\D}[1]{D[{#1}]}

\begin{document}

\preprint{APS/123-QED}

\title{Pseudogap formation in the moderate correlated layered attractive Hubbard model}
\begin{abstract}
We consider the layered attractive Hubbard model with a moderate interaction strength at quarter filling. The Green's function, self-energy, and density of states are calculated for relatively large clusters using the fluctuating local field method within the approximation of vanishing effective interaction. The emergence of a pseudogap is demonstrated for the cluster system, and it is shown that integration over the fluctuating field is equivalent to the summation of zero-mode ladder diagrams above critical temperature. For the large layered system, a pseudogap in the density of states is obtained within a cluster scheme that treats the coupling between layers as a static $U(1)$ symmetry-breaking field.
\end{abstract}

\author{A. V. Belov}
\email{belov.aleksandr@phystech.edu}
\affiliation{Russian Quantum Center, Moscow 121205, Russia}
\affiliation{National University of Science and Technology “MISIS”, Moscow 119049, Russia}
\affiliation{Department of Physics, National Research University Higher School of Economics, Moscow, 101000, Russia}
\author{A. N. Rubtsov}
\affiliation{Russian Quantum Center, Moscow 121205, Russia}
\affiliation{Lomonosov Moscow State University, Moscow 119991, Russia}
\author{B. Krippa}
\affiliation{London School of Economics and Political Science WC2A 2AE, London, UK}

\date{\today}

\maketitle

\section{\label{sec:level1} Introduction}

Since the late 20th century, our understanding of superconductivity has undergone a significant paradigm shift driven by the discovery of superconducting systems that do not fit into the standard Bardeen–Cooper–Schrieffer (BCS) theory \cite{Bardeen1957}. While the mean-field BCS framework, built upon the concept of the formation of isotropic Cooper pairs below the critical temperature, successfully explains the behavior of conventional superconductors, it fails to capture the physics of strongly correlated and unconventional systems. In this work, we do not address the complex physics of high-temperature superconducting (HTSC) cuprates, where the pairing mechanism is governed by the unconventional $d$-wave symmetry arising from electronic correlations in layered materials \cite{Tsuei2000pairing, keimer2015quantum,lee2014amperean}. We also exclude iron-based superconductors, which exhibit a characteristic $s_{\pm}$-pairing symmetry resulting from the repulsive interband interaction between electron and hole pockets within a multiband Fermi surface \cite{Yu2013spin,Hirschfeld2011gap,chubukov2015iron}. Instead, we restrict our analysis exclusively to the single-component systems characterized by conventional $s$-wave pairing symmetry, where the order parameter retains a constant phase and the Fermi surface remains relatively simple.

The BCS limit formally corresponds to a weak attractive interaction between electrons, characterized by a small positive $s$-wave scattering length $a > 0$, such that $k_F a \ll 1$, where $k_F$ is the Fermi wave vector \cite{Randeria1995}. In this weak-coupling regime, Cooper pairs are loosely bound and their size greatly exceeds the average interparticle distance. There are two well-established experimental markers that signal a departure from this strict BCS limit. The first is an enhancement of the ratio of the zero-temperature spectroscopic gap $\Delta_0$ to the critical temperature $T_c$ beyond the universal BCS mean-field value of $\frac{2\Delta_0} {T_c} \approx 3.53$ \cite{Randeria1995} (hereafter we set Boltzmann constant to 1). A classic example is elemental lead (Pb), for which this ratio exceeds $4.2$, reflecting the onset of strong-coupling corrections captured by the Eliashberg theory \cite{Carbotte1990}. The second, more dramatic marker is the emergence of a pseudogap in the electronic spectrum at temperatures modestly above $\Theta_c$. This phenomenon signals the existence of preformed electron-hole pairs (or strong pairing fluctuations) already in the normal metallic phase, where the spectral weight of a single-particle is partially suppressed without the establishment of long-range phase coherence \cite{Chen2005}. The pseudogap is particularly pronounced in layered quasi-two-dimensional systems and in materials where the development of a superconducting phase is substantially suppressed by fluctuations \cite{Feld2011}. For definiteness, we note that the picture outlined above describes moderately correlated systems, in which the size of a Cooper pair still significantly exceeds the average distance between electrons. A further increase in the strength of the attractive interaction drives the system across the BCS--BEC crossover, where Cooper pairs evolve into compact tightly bound dimers, and the magnitude of the spectral gap becomes determined by their binding energy \cite{Randeria1995,Nozieres1985, NobelReview2003}.

A standard theoretical playground for studying the correlated s-wave superconductors is the attractive Hubbard model. The strength of electronic correlations is controlled by the ratio of the on-site Hubbard attraction $|U|$ to the nearest-neighbor hopping integral $t$. Moderate correlations are realized for $|U|/t \approx 1\ldots 2$, whereas larger values of this ratio correspond to the regime of strong correlations. In Ref. \cite{Gunnarsson2015fluctuation} it was shown from analysis of the Dyson-Schwinger equation in the spin, particle-particle, and charge channel that the general contribution to self-energy $\Sigma_k$ comes from the collective mode with momentum $\textbf{Q} = (0, 0)$ in the particle-particle channel. In other words, the superconducting instability becomes the dominant channel when the system is doped away from half-filling, where charge-density-wave and superconducting correlations are degenerate at the particle-hole symmetric point. Parquet decomposition \cite{Astretsov2020dual, Gunnarson2016parquet} is another approach in which the irreducible two-particle vertex are decomposed in all possible fluctuation channels and, thus, we do not require an information about the dominant channel of fluctuations. However, this method suffers from intrinsic divergence in the regime of parameters where the pseudogap physics is arose.

The emergence of a pseudogap in the attractive Hubbard model has been observed in quantum Monte Carlo (QMC) simulations \cite{Moreo1992, singer1996bcs, singer1999spectral}. It was shown that the spectral weight in the density of states pushed away from the Fermi level and redistributed over the entire band, rather than only around the gap, as the temperature decreased. It means that strong correlations in this system are predominantly local in character, which significantly simplifies the problem of their theoretical description. In particular, the use of dynamical mean-field theory (DMFT) becomes well justified in this case. Such an investigation was carried out in Ref.~\cite{Peters2015local}, where it was explicitly demonstrated. The local approximation correctly captures the formation of a pseudogap, which persists in the system at temperatures significantly exceeding the superconducting critical temperature and at large interaction values for 3D case. 

At the same time, analytical and semi-analytical methods for describing of the pseudogap in the regime of moderate correlations \cite{Allen2001nonpertubative, Kyung2001pairing, Gauvin-Ndiaye2023improved, Martin2023Nonlocal}, where pairing fluctuations acquire an essentially nonlocal nature, are far less developed. The difficulty lies in constructing a fundamentally nonlocal theory that remains applicable over a wide temperature range, including the critical region in the vicinity of the phase transition. The onset of the pseudogap upon lowering the temperature from the high-temperature side can be understood as a result of summing of the specific diagrammatic series. In particular, the self-consistent theory based on the summation of ladder diagrams within the particle--particle channel provides a quantitatively accurate description of the pseudogap formation at temperatures moderately above $T_c$ \cite{Kyung2001pairing}. In this approach result of the ladder summation is used in Dyson-Schwinger equation and simple renormalization of interaction vertex. However, as the temperature is lowered further towards the critical point, these approaches break down. Formally, this failure signals the necessity to account for higher-order diagrams, most notably two-particle vertex corrections and diagrams with crossing interaction lines, which becomes technically prohibitive in practice.

In this work, we study the layered attractive Hubbard model with $U=2 t$ at the quarter filling, using the fluctuating local field (FLF) method \cite{rubtsov2018fluctuating, lyakhova2024fluctuating}. This method reduces the initial problem on a finite lattice to a weighted ensemble of mean-field problems. Our approach accounts for significant non-local fluctuations in the dominant particle-particle channel. We then extend the resulting self-energy to the infinite system.
As a result, we observe a clear pseudogap in the density of states. The intermediate-coupling regime is indicated by a renormalization of the spectral weight, which occurs in a narrow region near the pseudogap.
The paper is structured as follows.
In the Methods section, we present the cluster scheme and the fluctuating local field formalism. The Results section presents the density of states for both the cluster and the infinite lattice, along with cluster order parameter dependence on the temperature. Finally, in the Discussion section, we summarize and interpret our results. 

\section{\label{methods} Method}
\begin{figure}
    \centering
    \includegraphics[width=0.99\linewidth]{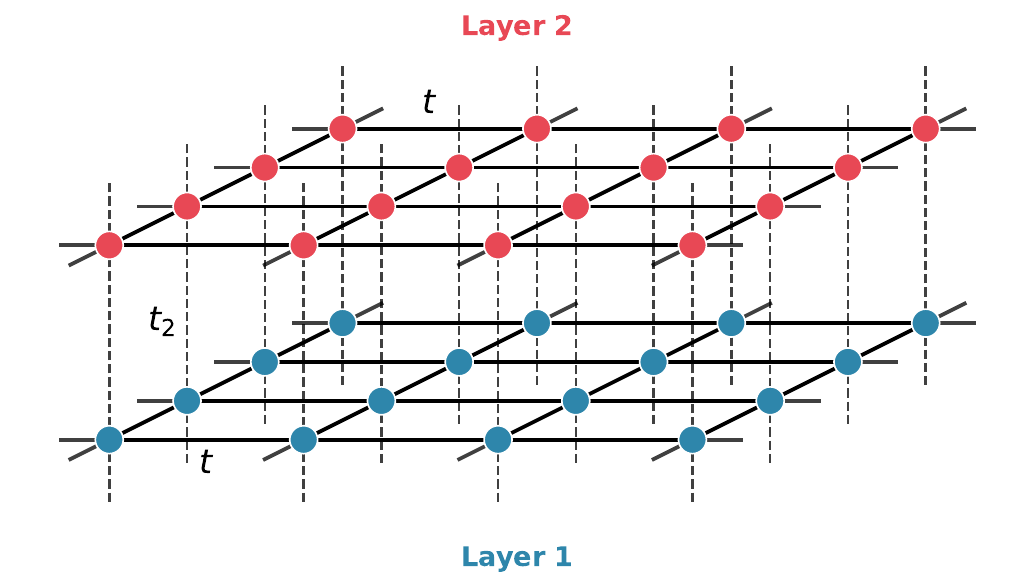}
    \caption{Schematic illustration of the layered system. Black solid lines denote intralayer hopping, while grey dotted lines denote interlayer hopping.}
    \label{fig:schema_layer}
\end{figure}

We consider the layered attractive Hubbard model (see Fig. \ref{fig:schema_layer}) with the Hamiltonian
\begin{equation}
    \begin{split}
    H = \sum_{\av{rr'}\sigma} t_{rr'} c^{\dag}_{r\sigma} c_{r'\sigma}
    - \mu \sum_{r\sigma} n_{r\sigma} \\
    - U \sum_{r} \left(n_{r\uparrow} - \frac{\av{n}}{2}\right)
    \left(n_{r\downarrow} - \frac{\av{n}}{2}\right),
    \end{split}
\end{equation}
where $c_{r\sigma}(c^{\dagger}_{r\sigma})$ are fermionic annihilation (creation) operators at site $r$ of the lattice with spin $\sigma$, $n_{r\sigma} = c^{\dagger}_{r\sigma} c_{r\sigma}$ is the electron density operator, and $\av{n} = \av{n_{\uparrow}} + \av{n_{\downarrow}}$ is the average electron density. The parameter $U$ denotes the interaction strength.

The notation $\av{rr'}$ denotes a summation over the nearest-neighbour sites. We set $t_{rr'} = -t$ for the hopping between the nearest neighbours within the same layer and $t_{rr'} = -t_2$ for the hopping between neighbouring layers. In the latter case, the electron can hop only between the sites located directly above or below each other in an adjacent layers. We assume that $t_2$ is relatively small, i.e. the interlayer coupling is weak. Periodic boundary conditions are imposed within each layer.

We study this system for two reasons. First, in the purely two-dimensional case, a true long-range order is destroyed by thermal fluctuations according to the Mermin--Wagner theorem \cite{Mermin1966Absense}. Second, we expect that the Berezinskii--Kosterlitz--Thouless (BKT) physics \cite{berezinskii1972destruction, kosterlitz1973ordering} does not significantly influence the pseudogap behaviour in the layered system, which allows us to focus primarily on the effects of the superconducting fluctuations.

We reduce a lattice problem to the effective $L\times L$ cluster problem in the presence of a static $U(1)$ symmetry-breaking field $\Delta$ using the cluster scheme discussed below. Unlike in other cluster schemes, we assume that $\Delta$ does not depend on frequency. It is known that a frequency-dependent hybridization field $\Delta(\omega)$ improves convergence \cite{maier2005quantum}; however, the advantage of the FLF method is the possibility to perform  calculations for relatively large clusters. Therefore, the static approximation provides sufficiently good convergence in our case. The Hamiltonian of the cluster problem in the external field is
\begin{equation}
   \begin{split}
    H_{\text{cluster}} =
    -t\sum_{\av{ij}\sigma}
    c^{\dag}_{i\sigma} c_{j\sigma}
    - \mu \sum_{i\sigma} n_{i\sigma}
    + \Delta \sum_i c^{\dagger}_{i\uparrow} c^{\dagger}_{i\downarrow}
    + \text{h.c.} \\
    - U \sum_i
    \left(n_{i\uparrow} - \frac{\av{n}}{2}\right)
    \left(n_{i\downarrow} - \frac{\av{n}}{2}\right),
   \end{split}
\end{equation}
where $\av{ij}$ denotes summation over the nearest-neighbour sites within the same layer.

We calculate the cluster self-energy $\Sigma_K$ for $\Delta \ne 0$ and construct the self-energy $\Sigma_k$ of a large layer within the cluster scheme. Because of the weak inter-layer hopping we treat lattice as an effective background that creates field $\Delta$ in which we immerse cluster. Finally, we extract the pseudogap behavior from the density of states.

\subsection{Fluctuating local field}
Now we are in a position to discuss the method used to solve the cluster problem introduced above. Since we are interested in Green's function (GF), we begin the calculation of real-frequency observables within the FLF approach \cite{lyakhova2024fluctuating} by multiplying the partition function by the functional unity in the following form:

\begin{equation}
\begin{split}
    Z_{FLF} = \mathscr{N}\int d\text{Re}\, \nu d\text{Im}\, \nu \,
    e^{-\frac{\beta N}{2\lambda}(\nu + \lambda \theta - \Delta)\overline{(\nu + \lambda \theta - \Delta)}} \cdot \\ 
    \int \D{c^{\dag}, c} e^{-S_0 - S_{int}},
\end{split}
\end{equation}
where $\mathscr{N}$ is a normalization constant, $S_0$ is the free action of the cluster, and $S_{int}$ is the interacting part of the cluster action. $\nu = \text{Re}\, \nu + i\text{Im}\, \nu$ is the complex $U(1)$ symmetry-breaking field, the parameter $\lambda$ is a free parameter, and the cluster order parameter in the Cooper channel is defined as

\[
\theta = \frac{1}{\beta N}\int_0^{\beta} d\tau \sum_i 
c^{\dag}_{i\uparrow}(\tau)c^{\dag}_{i\downarrow}(\tau).
\]

Using spatial and imaginary-time uniformity of this cluster problem, after a straightforward algebra we obtain

\begin{equation}
\begin{split}
   Z_{FLF} = \mathscr{N}\int d\text{Re}\, \nu d\text{Im}\, \nu \,
    e^{-\frac{\beta N}{2\lambda}|\nu + \Delta|^2} \cdot \\
    \int \D{c^{\dag}, c}e^{-S_0(\nu)-S^{'}_{int}}, 
\end{split}
\end{equation}
where
\[
S_0(\nu) = S_0 + \nu\sum_kc^{\dagger}_{k\uparrow}c^{\dagger}_{-k\downarrow} + h.c.,
\]

\[
    \begin{split}
       S^{'}_{int} =
       \frac{\lambda}{\beta N}
       \sum_{k, q}
       c^{\dagger}_{k\uparrow}
       c^{\dagger}_{-k\downarrow}
       c_{-q\downarrow}
       c_{q\uparrow} \\
       - \frac{U}{\beta N}
       \sum_{k, k', q}
       c^{\dagger}_{k+q\uparrow}
       c^{\dagger}_{k'-q\downarrow}
       c_{k'\downarrow}
       c_{k\uparrow},
    \end{split}
\]
and $k=(i\omega_{k}, \vec{k})$ is the four-vector in frequency--momentum space. From the expression for $S'_{int}$ we see that for $\lambda = U / 2$ the contribution with $k=-k'$ corresponding to the attractive interaction is compensated by the effective repulsive interaction. Physically, this means that such a choice of the free parameter $\lambda$ removes the zero-mode divergence of the Cooper susceptibility within the mean-field approximation. In addition, the choice $\lambda=U/2$ guarantees that the mean-field equations define a saddle point of the integral over the fluctuating field.

To extract the spectral properties of the cluster and the lattice, we calculate the Green's function matrix \cite{maier2005quantum} within the FLF approach:

\begin{equation}
    \hat{G}^{FLF}_K =
    \begin{pmatrix}
        G_K & F_K \\
        F^{*}_{K^*} & -G_{-K}
    \end{pmatrix},
\end{equation}

where $K=(z, \vec{K})$ and $z$ is the complex-valued frequency. 

The zero-mode contribution in the attractive channel is compensated by effective repulsion, as discussed above. However, interactions in all other channels remain and therefore an additional approximation is required for further calculations. We assume that $S'_{int}\approx 0$ and call this approximation $\text{FLF-}0$. In this case, the GF for each value of the field $\nu$ reduces to the mean-field Green function:

\[
    \hat G^{(0)}_K(\nu) =
    \begin{pmatrix}
        z - \xi_{\vec{K}} & \nu \\
        \bar \nu & z + \xi_{\vec{K}}
    \end{pmatrix}^{-1}.
\]

For each problem with a fixed field value $\nu$, the system possesses a two-dimensional rotational symmetry. This allows us to integrate out the angular dependence explicitly, which leads to an effective weight depending only on the absolute value of $\nu$. Thus, the FLF Green functions in the zeroth-order approximation are
\begin{gather}
    G_K = \av{G^{(0)}_K(\nu)}_0, \\
    F_K = \av{F^{(0)}_K(\nu)}_1,
\end{gather}
where
\[
\av{...}_n =
\frac{1}{Z_{\text{FLF}}}
\int d|\nu|\, w_n(|\nu|, \Delta),
\]
and $w_n(|\nu|, \Delta)$ is the effective weight function. This weight function determines the contribution of each ensemble configuration:

\begin{equation}
  w_n(|\nu|, \Delta) =
  2\pi|\nu|
  e^{-S^{FLF}(|\nu|, \Delta)}
  I_n\left(\frac{\beta N}{U}2\Delta|\nu|\right),
\end{equation}
where $I_n(x)$ is the modified Bessel function of the first kind of order $n$, and $S^{FLF}(|\nu|, \Delta)$ is the effective FLF action:
\begin{equation}
    S^{FLF} =
     \frac{\beta N}{U}(|\nu|^2 + \Delta^2)
     -\sum_{\vec{K}}
     \left[
     \beta\lambda_{\vec{K}}
     + 2\ln{(1 + e^{-\beta \lambda_{\vec{K}}})}
     \right],
     \label{eq:effective_FLF_action}
\end{equation}
where $\lambda_{\vec{K}} = \sqrt{\xi_{\vec{K}}^2 + |\nu|^2}.$ The partition function within the FLF approach takes the form 
\[
    Z_{\text{FLF}} = \int d|\nu|w_0(|\nu|).
\]

Now we are in position to extract real-frequency properties of a system and calculate the self-energy. We take mean-field expression for the Green function and integrate out a dependence on the fluctuating field $\nu$ to find the retarded FLF Green function. The density of states is given by
\begin{equation}
    \rho(\omega) =
    -\frac{2}{\pi}
    \Im{\frac{1}{N}\sum_{\vec{K}}G_{\vec{K}}^R(\omega)}.
\end{equation}
 
We also perform calculations within first-order perturbation theory by including the Fock diagram originating from the effective repulsive interaction. This approximation we call $\text{FLF-}1$.The Fock contribution gives rise to the normal self-energy
\[
    \Sigma^{F}_K =
    \frac{U}{\beta N}G_{-K}.
\]

One can see that the choice $\lambda = U / 2$ also leads to a cancellation of the anomalous Hartree diagram. Therefore, within first-order perturbation theory only the Fock contribution remains. We add this self-energy correction to the Green's function averaged over the fluctuating field.

We work with the clusters with the sizes: $8\times8$ and $16\times16$. Such sizes allow us to choose a handy cluster scheme and claim that for a description of a relevant physics within the FLF approach we need only one collective mode with $\textbf{Q}=(0,0)$.

\subsection{Cluster scheme}
We are interested in the physics of a large lattice system, so we need some scheme that would allows us to use information about cluster and expand it to all lattice. Also, density of states of the cluster is a set of the delta-peaks which have high degeneracy, because of dimensionality of point-group symmetry of the square lattice. To solve this problem we immerse a cluster into a big lattice. This procedure can be implemented in a variety of ways say in the dynamical cluster approximation (DCA) \cite{maier2005quantum} for example. However, it is known that with increasing the cluster size the convergence of the chosen scheme becomes better and the difference between possible  schemes becomes smaller. Because of the sufficiently large size of the cluster (in our approach $16\times 16$) we use a simple case of a static field. Thus,  we treat the lattice as an effective background that creates the static  $U(1)$ symmetry-breaking field $\Delta$ in which we immerse the cluster. 

Earlier we note that the hopping between layers is small, i.e. $t_2 \ll U$. After calculating the effective hamiltonian, one can see that the inter-layer hopping leads to effective exchange interaction with the value $2t^2_2/U$. We treat this interaction in the mean-field approach that gives contribution $2t^2_2/U\cdot \sum_i \Delta c^{\dagger}_{i\uparrow}c^{\dagger}_{i\downarrow}$. Thus, we find external field $\Delta$ from the equation
\begin{equation}
    \Delta - \frac{2t^2_2}{U}\bar F^{cl}(\Delta) = 0,
    \label{eq:self_Delta}
\end{equation}
where $\bar F^{cl}(\Delta)$ is fully local anomalous Green's function in the FLF and its value is related to the order parameter:
\begin{equation}
    \bar F^{cl}(\Delta)=\av{\frac{1}{N}\sum_{\vec{K}}\frac{|\nu|}{2\lambda_{\vec{K}}}\tanh{\frac{\beta \lambda_{\vec{K}}}{2}}}_1,
\end{equation}

For finding the lattice DOS we first calculate the cluster self-energy $\Sigma_K$ and then, using a rather standard algorithm we calculate the self-energy of the infinite layer  $\Sigma_k$. Namely, we take the following steps: 
    \begin{enumerate}
        \item Find normal self-energy $\Sigma_K$ and anomalous self-energy $\Phi_K$ of the cluster using FLF from Dyson's equation:
        \[\hat \Sigma_K = [\hat{G}^{(0)}_K]^{-1} - [\hat{G}^{FLF}_K]^{-1},\]
        where 
        \[
            \hat \Sigma_K = \begin{pmatrix}
                \Sigma_K & \Phi_K \\
                \Phi^*_{K^*} & -\Sigma_{-K}
            \end{pmatrix}.
        \]
        \item To construct lattice self-energy from cluster by:
        \[\Sigma_k = \Sigma_K,\ \text{if}\ \vec{k} \in P(\vec{K}),\]
        where $P(\vec{K})$ is a square area in the Brillouin zone with center in $\vec{K}$;
        \item Define the lattice GF as 
        \[G^{R}_{\vec{k}}(\omega) = \frac{\zeta_h}{\zeta_h\zeta_p - [\Phi^{R}_{\vec{k}}(\omega)]^2},\]
        where $\zeta_h = \omega + i\delta + \xi_{\vec{k}} + \Sigma^A_{\vec{k}}(-\omega)$, $\zeta_p = \omega + i\delta - \xi_{\vec{k}} - \Sigma^R_{\vec{k}}(\omega)$.
    \end{enumerate}

\begin{figure*}[!htbp]
    \centering
    \begin{subfigure}{0.32\textwidth}
        \includegraphics[width=\linewidth]{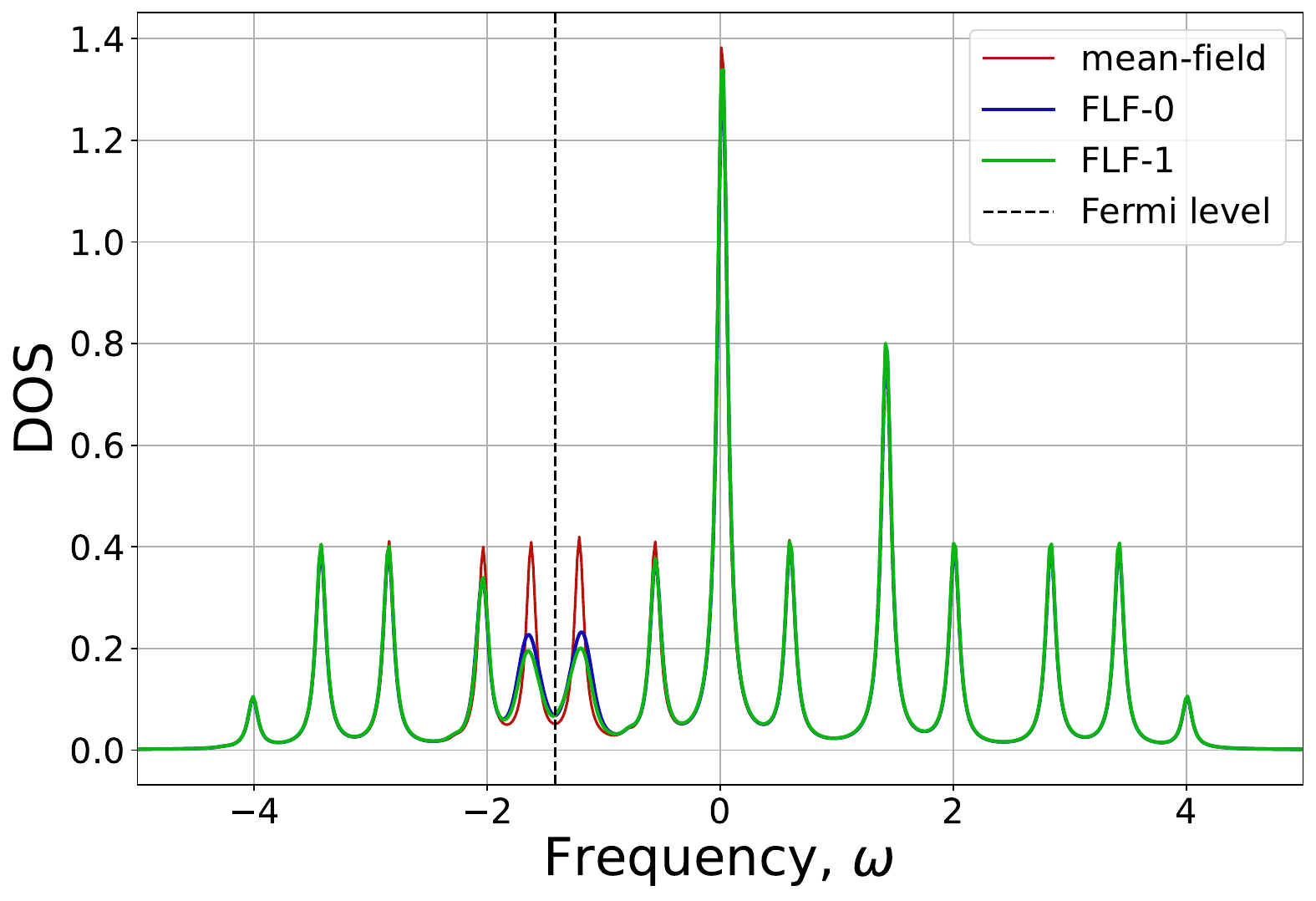}
        \caption{$8\times8,\ T=0.10$}
        \label{fig:8x8_below_Tc}
    \end{subfigure}
    \begin{subfigure}{0.32\textwidth}
        \includegraphics[width=\linewidth]{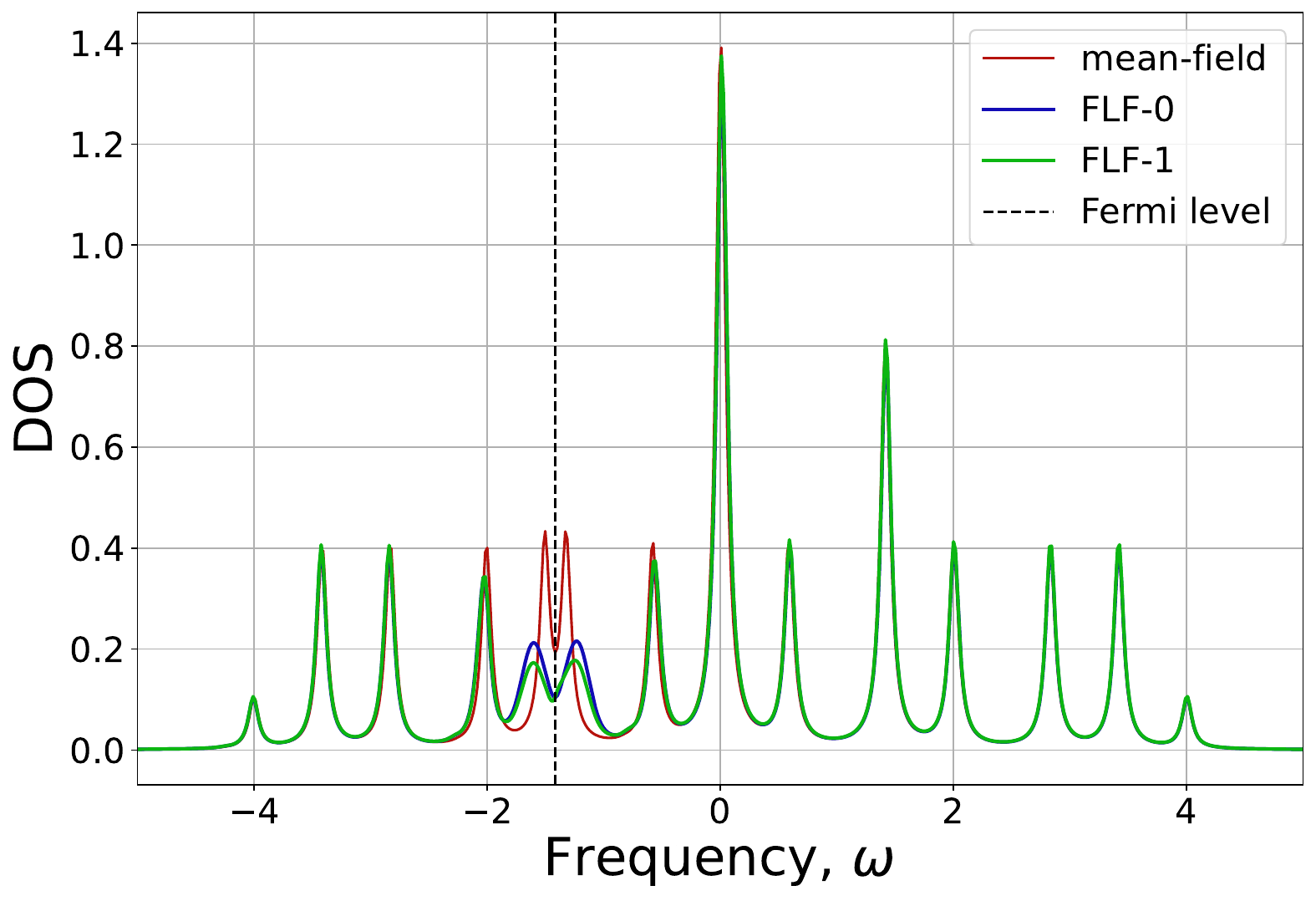}
        \caption{$8\times8,\ T=0.13$}
        \label{fig:8x8_near_Tc}
    \end{subfigure}
    \begin{subfigure}{0.32\textwidth}
        \includegraphics[width=\linewidth]{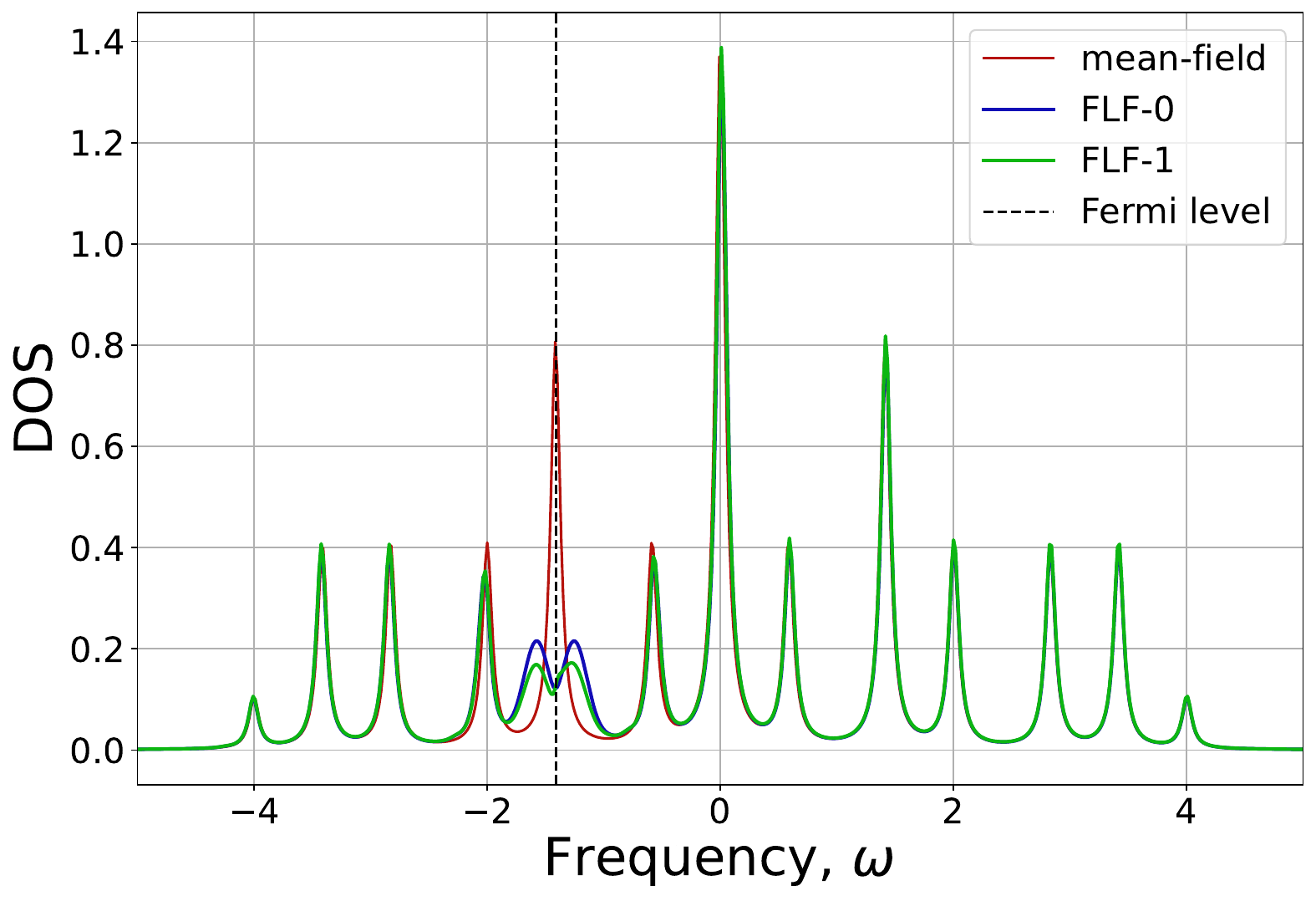}
        \caption{$8\times8,\ T=0.15$}
        \label{fig:8x8_above_Tc}
    \end{subfigure}

    \vspace{0.3cm}

    \begin{subfigure}{0.32\textwidth}
        \includegraphics[width=\linewidth]{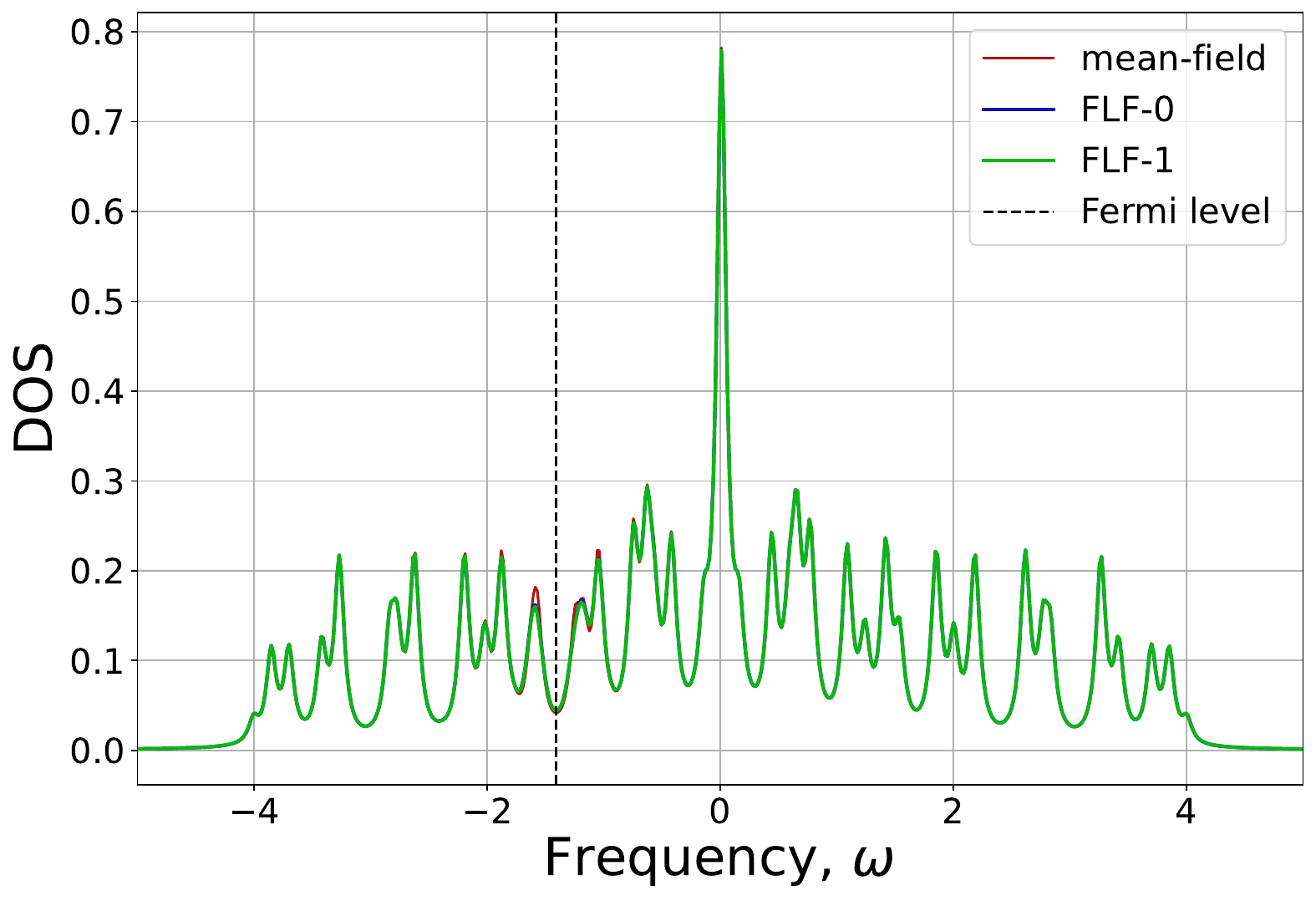}
        \caption{$16\times16, \ T=0.06$}
        \label{fig:16x16_below_Tc}
    \end{subfigure}
    \begin{subfigure}{0.32\textwidth}
        \includegraphics[width=\linewidth]{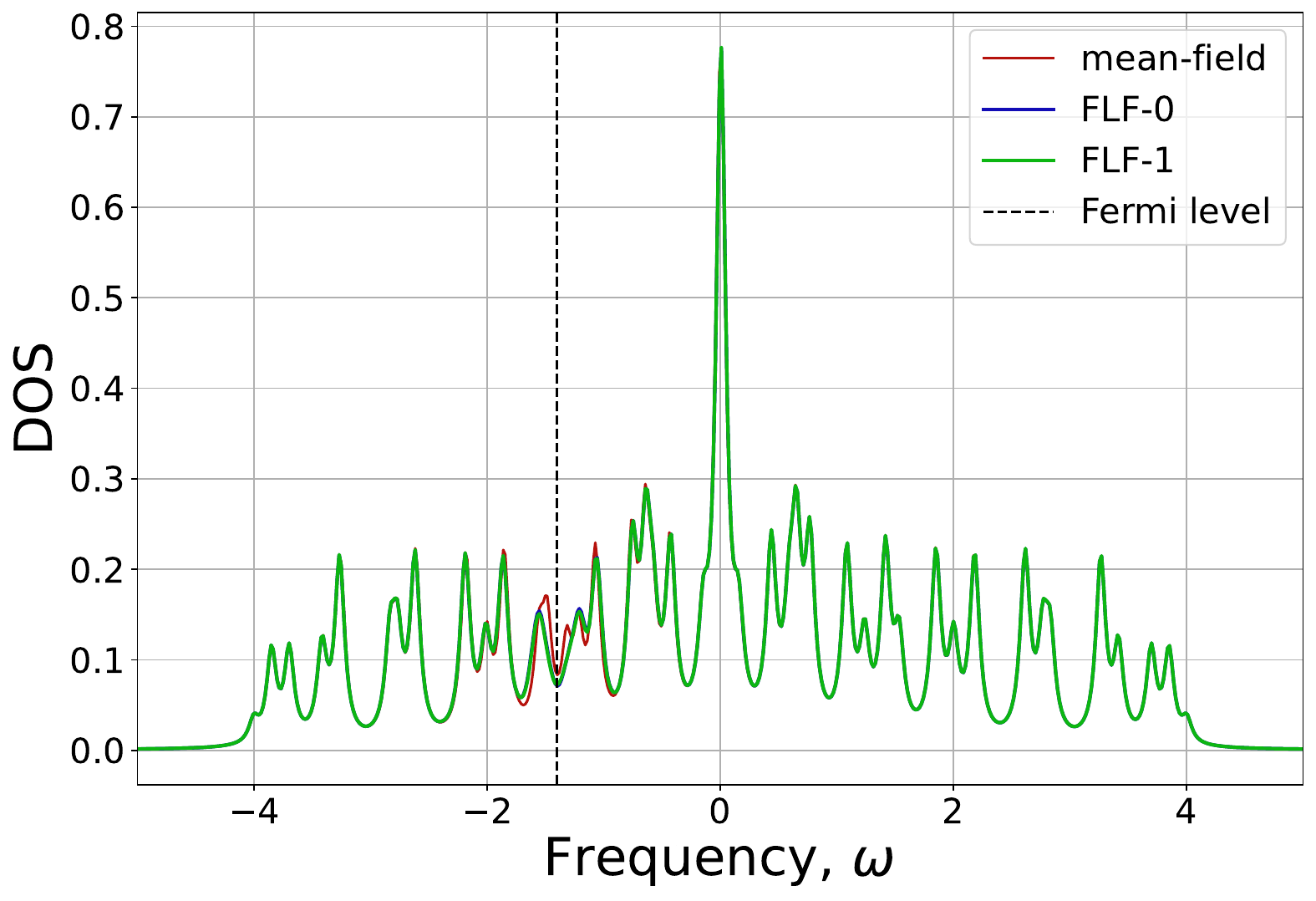}
        \caption{$16\times16, \ T=0.09$}
        \label{fig:16x16_near_Tc}
    \end{subfigure}
    \begin{subfigure}{0.32\textwidth}
        \includegraphics[width=\linewidth]{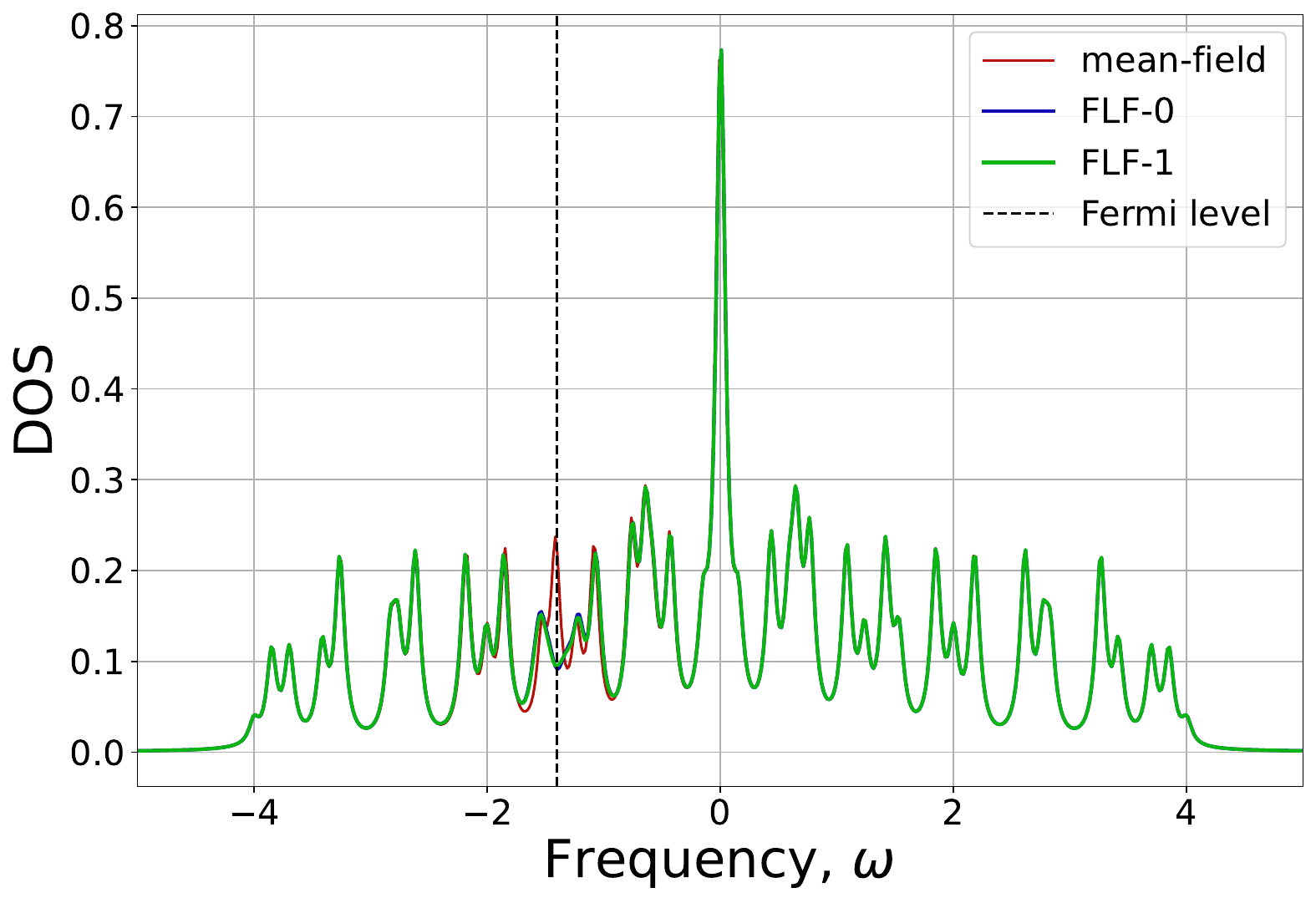}
        \caption{$16\times16, \ T=0.12$}
        \label{fig:16x16_above_Tc}
    \end{subfigure}

    \caption{Density of states (DOS) for temperatures below, near and above mean-field transition for $8\times8$ (first row) with $T_c\approx 0.14$ and $16\times 16$ (second row) with $T_c\approx 0.1$. Red lines is mean-field DOS, blue - $\text{FLF-}0$ DOS, green - $\text{FLF-}1$ DOS, black slashed line is Fermi level. At this line, formation of a gap and a pseudogap is clearly seen. DOS from $\text{FLF-}1$ with Fock correction is nearly the same as DOS in $\text{FLF-}0$}
    \label{fig:DOS_for_diff_sizes}
\end{figure*}

\section{Results}
In this section, we first discuss the results of our approach for the cluster density of states, $\rho(\omega)$, for $\Delta = 0$. We then discuss the density of states in a large lattice based on the cluster solution. We consider clusters of sizes $16\times16$ and $8\times8$ at $\langle n_{\uparrow}\rangle = \langle n_{\downarrow}\rangle \approx 0.25$ (quarter filling), with $t=1$ and $U=2t$, at different temperatures and with periodic boundary conditions.

The mean-field problem for the fluctuating-field ensemble is free-fermionic. Therefore, we need to choose the chemical potential carefully because of the high symmetry of the square lattice. This symmetry gives rise to a relatively sparse set of peaks in the density of states (DOS), while the DOS vanishes between them. However, this is not the gap of interest. Instead, we aim to observe the formation of a gap and a pseudogap at the Fermi level. Therefore, we choose the chemical potential $\mu$ such that it coincides with one of the peaks, corresponding to $\langle n\rangle \approx 0.5$.

We study the cluster density of states in three regimes: below, near, and above the mean-field transition temperature $T_c$. In Figs.~\ref{fig:8x8_below_Tc} and~\ref{fig:16x16_below_Tc}, we compare the density of states obtained within the bare FLF approach ($\mathrm{FLF}\text{-}0$), the FLF approach with the Fock correction ($\mathrm{FLF}\text{-}1$) and the mean-field approximation. In all regimes, we clearly see that the Fock correction does not lead to a significant change in the calculated DOS. We regard this observation as evidence of the good convergence of the FLF scheme, suggesting that higher-order corrections are also expected to be small. Therefore, in what follows, we focus on the $\mathrm{FLF}\text{-}0$ results.  

Let us now analyse the FLF results for the cluster density of states in more detail. The first important observation is that the DOS changes only within a relatively narrow energy window around the Fermi level, indicated by the dashed vertical line in Fig.~\ref{fig:DOS_for_diff_sizes}. As discussed in the introduction, this behavior is typical of weakly and moderately correlated systems, where the correlations are predominantly non-local. Well below $T_c$, the FLF result is very close to the mean-field one, as shown in Figs.~\ref{fig:8x8_below_Tc} and \ref{fig:16x16_below_Tc}. This is expected because fluctuations are almost absent in this regime, so the FLF integral is essentially determined by its saddle-point value. In contrast, Figs.~\ref{fig:8x8_near_Tc} and \ref{fig:16x16_near_Tc} show that although the mean-field gap closes at the transition temperature, it persists in the FLF results. This behavior is interpreted as the formation of a pseudogap induced by superconducting fluctuations.

\begin{figure*}[!htbp]
    \centering

    \begin{subfigure}{0.32\textwidth}
        \includegraphics[width=\linewidth]{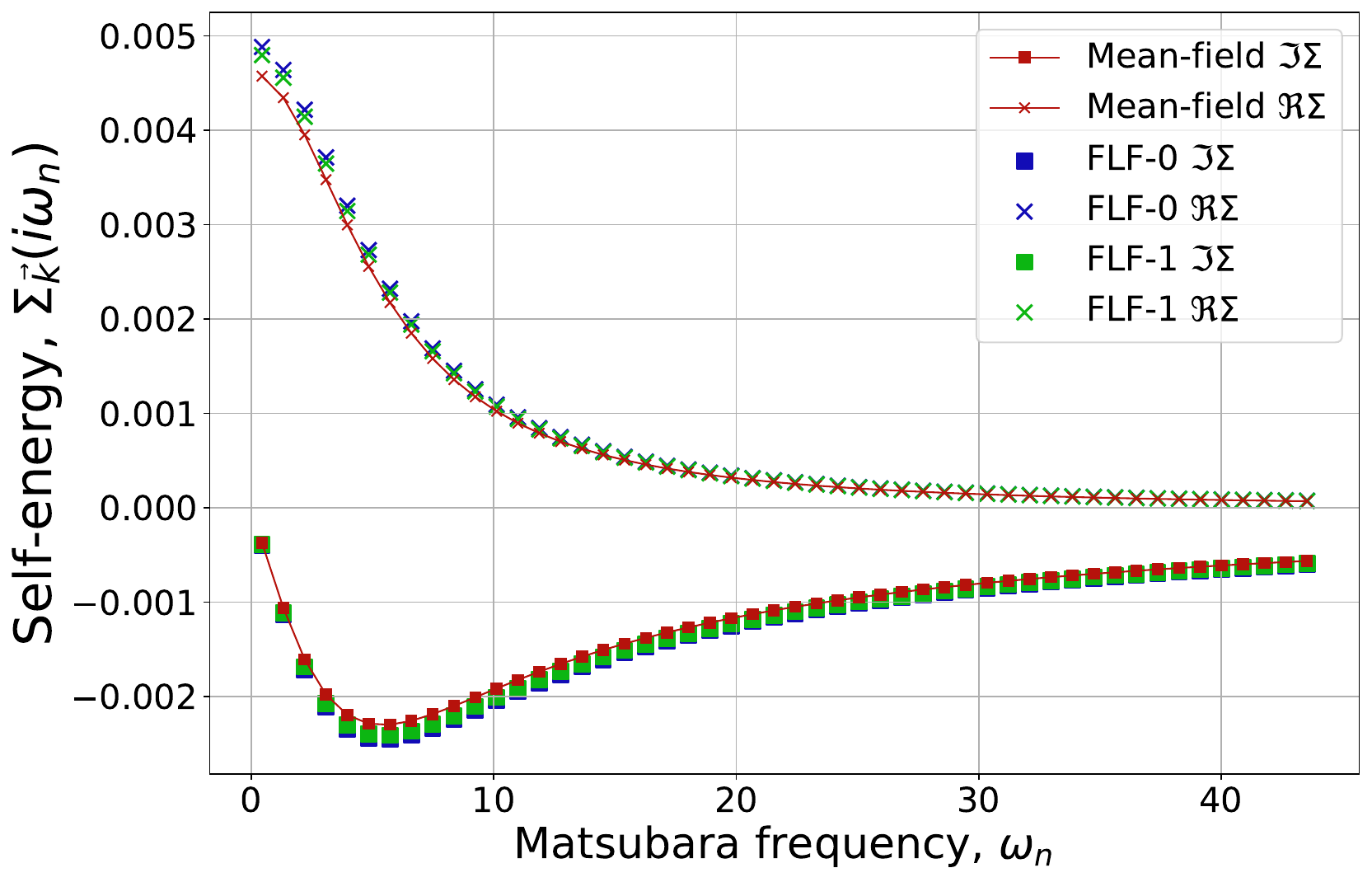}
        \caption{$T=0.05$}
        \label{fig:self_energy_below_Tc}
    \end{subfigure}
    \begin{subfigure}{0.32\textwidth}
        \includegraphics[width=\linewidth]{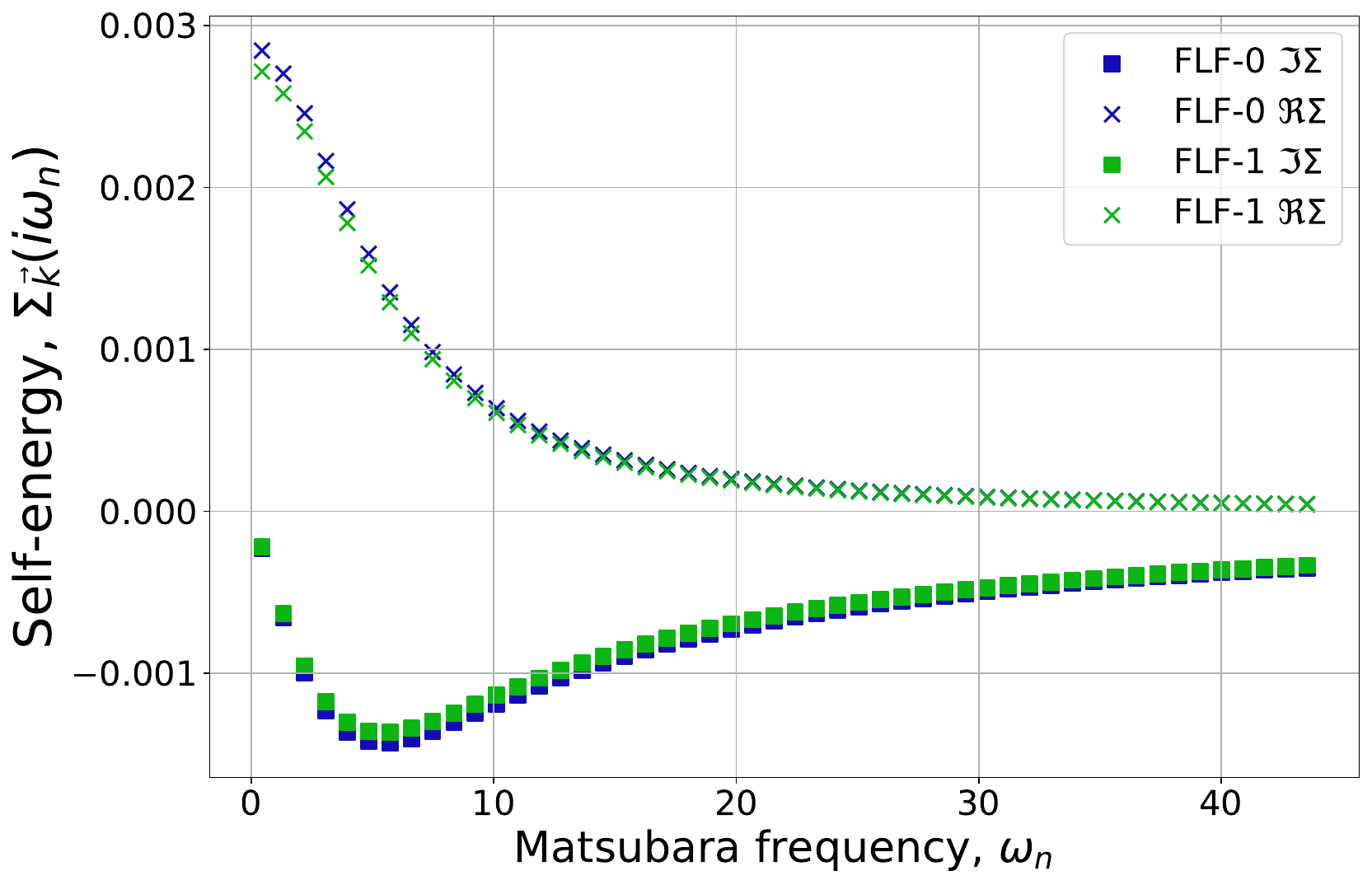}
        \caption{$T=0.09$}
        \label{fig:self_energy_near_Tc}
    \end{subfigure}
    \begin{subfigure}{0.32\textwidth}
        \includegraphics[width=\linewidth]{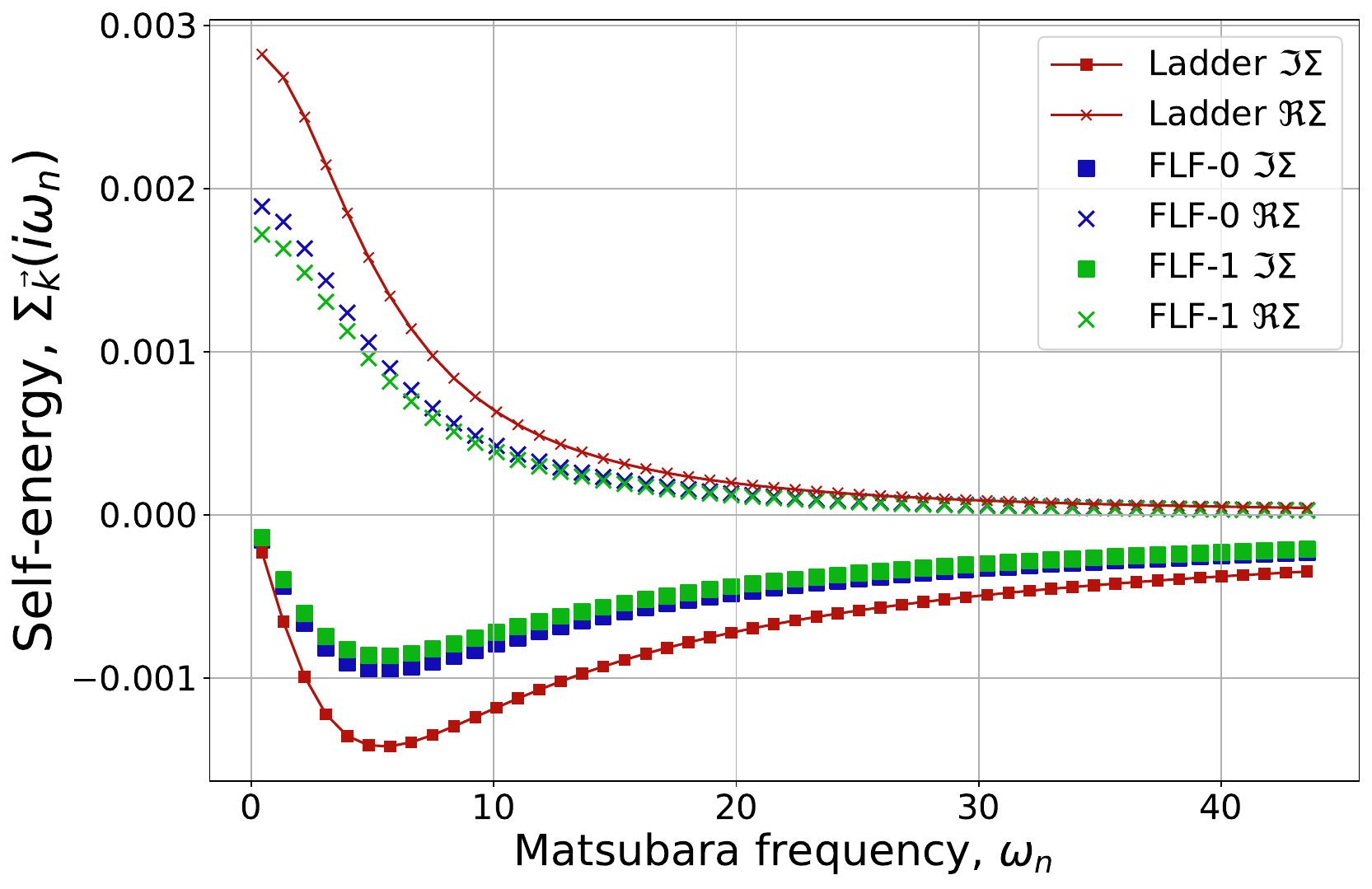}
        \caption{$T=0.12$}
        \label{fig:self_energy_above_Tc}
    \end{subfigure}

    \caption{Self-energy in FLF method for tempetoratures below, near and above mean-field transition for $16\times 16$ cluster with $T_c\approx 0.1$. In (a) FLF self-energy is compared with mean-field - red color, in (c) - is compared with zero-mode ladder approximation - also red color. Square markers denote imaginary part of self-energy, cross markers - real part}
    \label{fig:FLF_self_energy}
\end{figure*}

To gain insight into the physical origin of the pseudogap, we consider the self-energy at momentum $\vec{k}=(\pi,\pi)$. This quantity is plotted in Fig.~\ref{fig:FLF_self_energy} for the $16\times16$ cluster; the three panels correspond to the three temperatures introduced above. In Fig.~\ref{fig:self_energy_below_Tc} we again observe that well below $T_c$, the mean-field approximation reproduces the FLF results. The opposite limit, where the temperature is much higher than $T_c$, can also be treated perturbatively. As shown in Fig.~\ref{fig:self_energy_above_Tc}, the FLF results in this regime are reproduced by summing the ladder diagrams in the $\vec{k}=0$, $\Omega=0$ channel. This is fully consistent with the expected picture of weak fluctuations dominated by a single leading channel. However, near $T_c$  fluctuations become strong, and we are not aware of a perturbative approach that can be compared with the FLF results in the pseudogap regime. Therefore, we plot only the FLF self-energy. As shown in Fig.~\ref{fig:self_energy_near_Tc}, it remains well behaved, i.e., it does not exhibit any divergences.

\begin{figure}[!htbp]
    \includegraphics[width=0.99\linewidth, scale=1]{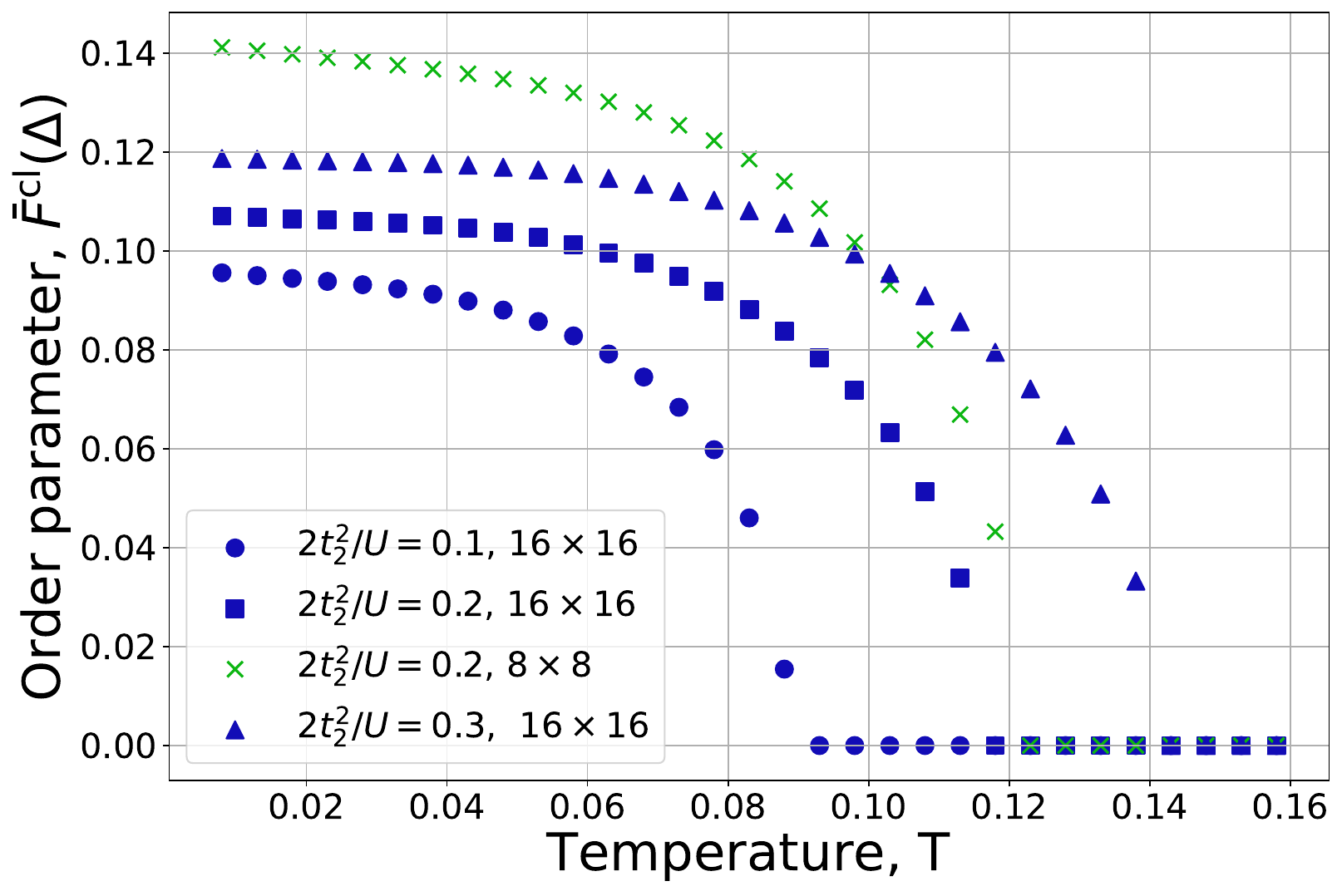}
    \caption{Dependence of order parameter on temperature with different exchange couplings for $16\times 16$ cluster. Pink circles - $0.1$,  violet triangles - $0.2$, green squares - $0.3$ and green crosses - $0.3$ for $8\times 8$ cluster}
    \label{fig:order_paremeter_diff_r}
\end{figure}

Let us now turn to the results for a large system. We employ the procedure described in the previous section. First, an effective field $2\Delta t_2^2/U$ is introduced to describe the interlayer coupling, and the order parameter $\Delta$ is determined self-consistently. Fig.~\ref{fig:order_paremeter_diff_r} shows the temperature dependence of the order parameter for the $16\times16$ cluster obtained for different values of the coupling parameter $2t_2^2/U$. For $2t_2^2/U = 0.2$, the results for the $8\times8$ cluster are also shown for a comparison. As one could expect, all the temperature dependencies are qualitatively similar, while the quantitative changes with the cluster size and the value of the coupling parameter remain relatively small.

\begin{figure*}[!htbp]
	\centering
	\begin{subfigure}[b]{0.32\textwidth}
        \centering
		\includegraphics[width=\linewidth]{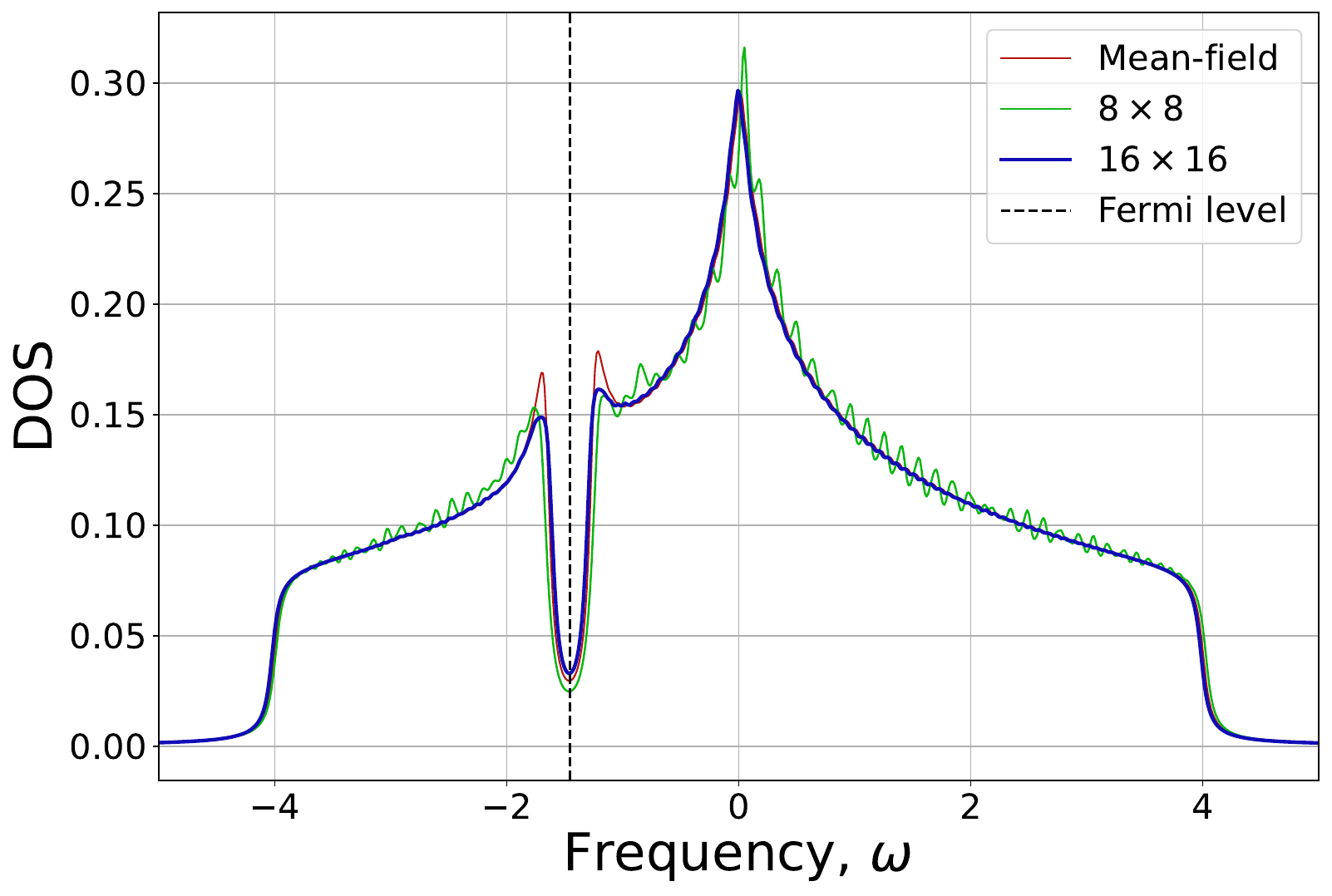}
		\caption{$T=0.08$}
		\label{fig:lattice_DOS_below_Tc}
	\end{subfigure}
    \hfill
	\begin{subfigure}[b]{0.31\textwidth}
        \centering
		\includegraphics[width=\linewidth]{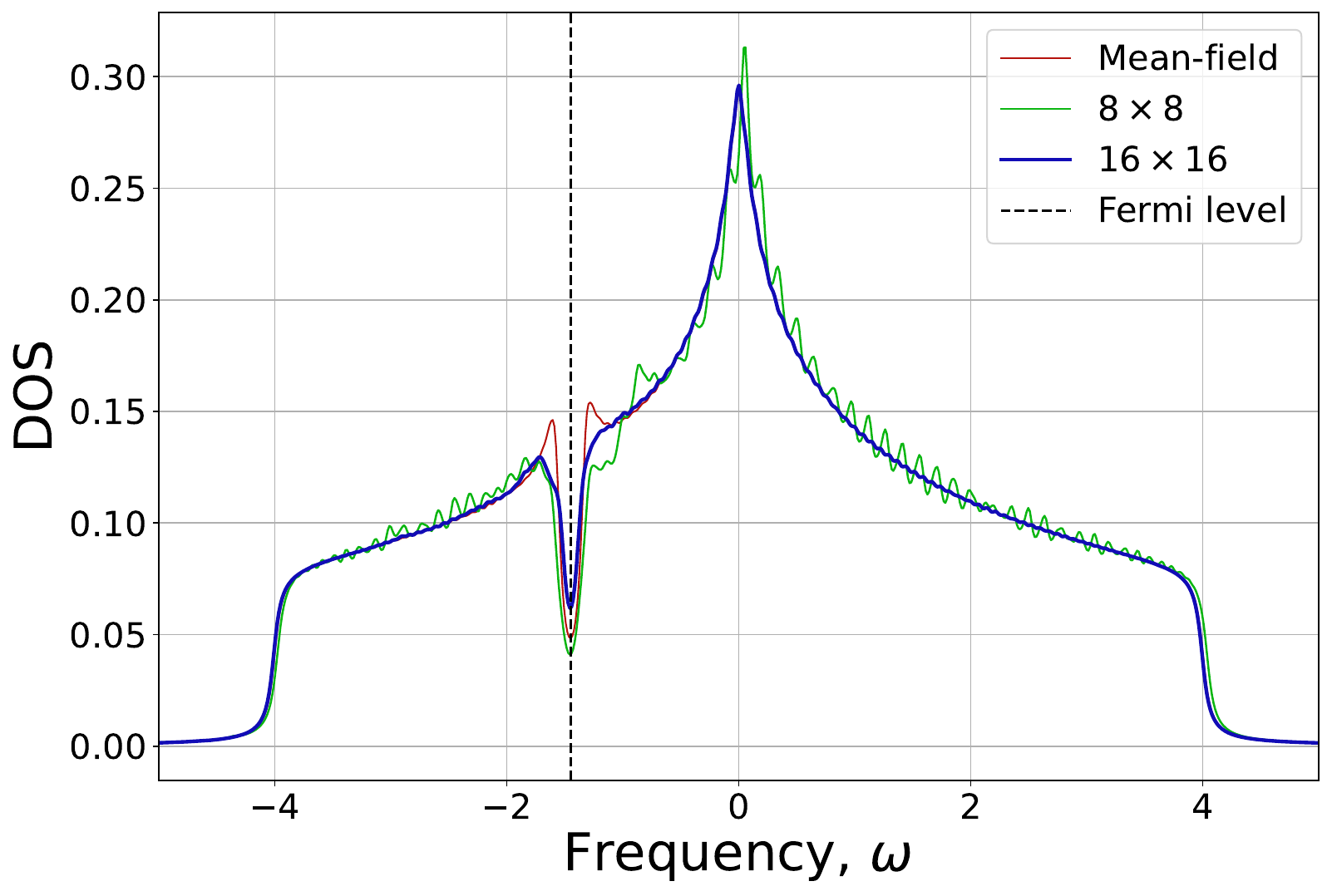}
        \caption{$T=0.11$}
		\label{fig:lattice_DOS_near_Tc}
	\end{subfigure}
    \hfill
        \begin{subfigure}[b]{0.32\textwidth}
        \centering
		\includegraphics[width=\linewidth]{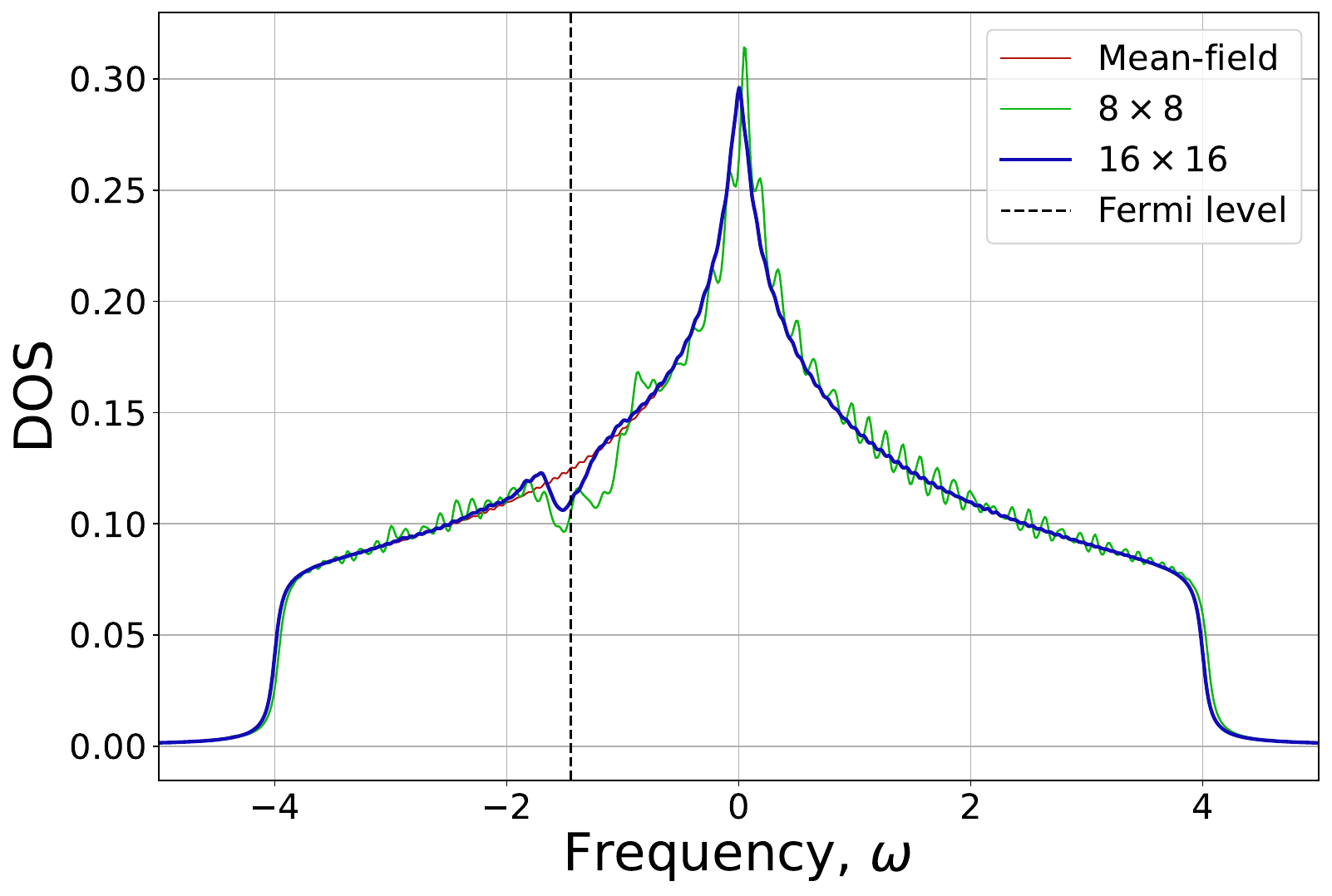}
		\caption{$T=0.14$}
		\label{fig:lattice_DOS_above_Tc}
	\end{subfigure}
	\caption{Density of states for large lattice near quarter-filling from discussed cluster scheme. Transition temperature from mean-field $T_c \approx 0.113$. Violet lines are DOS on base of the cluster with sizes $8\times8$, green lines - cluster with sizes $16\times 16$, pink lines - mean-field result, black dashed line - Fermi-level}
	\label{fig:lattice DOS}
\end{figure*}

After the self-consistent field has been determined, we extend the cluster self-energy to the infinite layer using the procedure described in the previous section. For $2t_2^2/U = 0.2$, this yields the density of states shown in Fig.~\ref{fig:lattice DOS}. In this figure, we compare the density of states for the infinite lattice obtained from the FLF solution of the cluster problem for the cluster sizes $8\times8$ and $16\times16$ with the mean-field results for the infinite planar lattice. In the calculations, we used the lattices of up to $240\times240$ sites and verified that a further increase of the lattice size did not affect the results.

\begin{figure}[!htbp]
    \centering
    \includegraphics[width=0.99\linewidth]{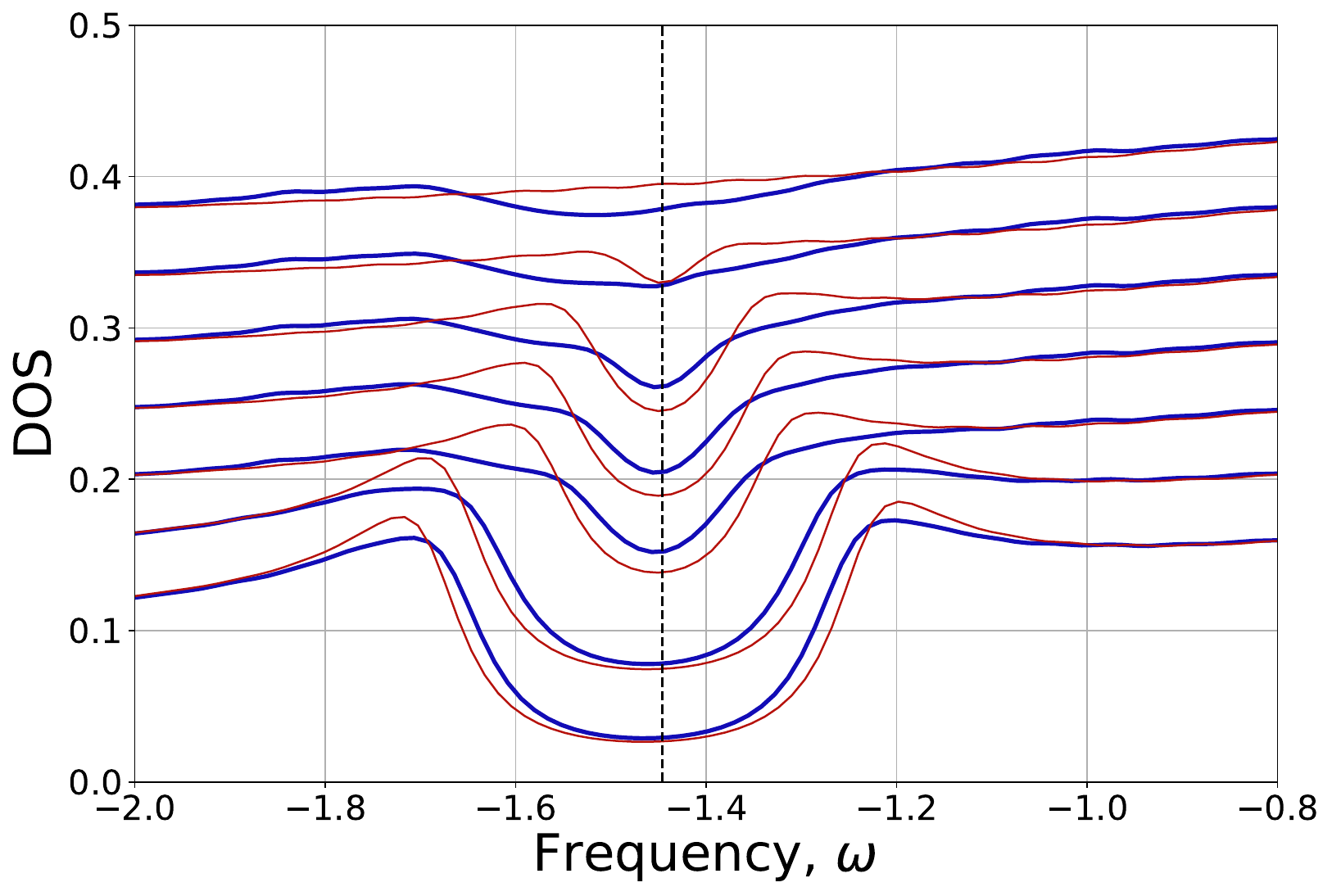}
    \caption{Density of states for large lattice near quarter-filling around Fermi level. Blue lines are DOS from cluster scheme on base of the $16\times 16$ cluster and red thin lines -- mean-field for large lattice, black dashed line -- Fermi level. From bottom to top line temperature is changed in following order: $0.05, \ 0.08, \ 0.11,\ 0.112,\ 0.114,\ 0.116,\ 0.12$}
    \label{fig:lattice_DOS_near_Fermi}
\end{figure}

Likewise Fig.~\ref{fig:16x16_below_Tc}, we observe that below $T_c$ the mean-field result is close to the FLF cluster data. In Fig.~\ref{fig:lattice_DOS_above_Tc} one can see that a pseudogap persists in the FLF density of states at $T=0.14$ for the lattice, whereas the mean-field gap is already closed. By comparing the obtained DOS using self-energies of the $8\times8$ and $16\times16$ clusters, we conclude that the finite cluster size gives rise to oscillatory features in the DOS. This artefact is clearly visible for the $8\times8$ cluster but is almost absent for the $16\times16$ cluster. We therefore conclude that a $16\times16$ cluster is sufficient to construct the lattice DOS. In Fig.~\ref{fig:lattice_DOS_near_Fermi} we can see in more detail gap-closing in mean-field and formation of a pseudogap in FLF-approach. In the next section we discuss shape of curves of DOS near Fermi level from FLF.

\section{Discussion}

We can now discuss the origin of the pseudogap formation in the layered attractive Hubbard model in more detail. As explained in the Methods section, the key idea of the FLF approach is the introduction of an alternative effective cluster description in which the collective mode at $\vec{K}=(0,0)$ and $\Omega=0$, responsible for the instability of susceptibility in the particle-particle channel, is represented by an integral over a fluctuating field. In this case, the partition function is multiplied by a functional unity. As a result, one obtains an ensemble of the mean-field problems in which zero mode is excluded from the interaction. This allows one to expect that the higher-order corrections in the perturbation theory are small. Indeed, it was shown in Ref.~\cite{Gunnarsson2015fluctuation} that the dominant contribution to the self-energy originates from the zero mode, accounting for more than half of the total contribution from the entire diagrammatic series. Therefore,  treating a zero mode exactly within the FLF formalism, one may expect that the remaining contributions have only a minor quantitative effect. We then assume that $S'_{\mathrm{int}}\approx 0$ and therefore work with purely mean-field problems. In contrast, the FLF-based cluster scheme allows one to describe the collective fluctuations of the infinite lattice in the vicinity of $\vec{k}=(0,0)$. Such a non-perturbative treatment of superconducting fluctuations leads to a modification of the density of states in the vicinity of the Fermi level.

To gain further insight into the nature of the pseudogap, we now turn to Fig.~\ref{fig:lattice_DOS_near_Fermi}, which shows the pseudogap formation within the FLF approach together with the gap closing in the mean-field approximation in a narrow energy window around the Fermi level. As can be seen from Fig.~\ref{fig:lattice_DOS_near_Fermi}, at sufficiently low temperatures the gaps obtained within the mean-field and FLF approaches are nearly identical. This is expected because, at low temperatures, the FLF and mean-field approaches become equivalent and therefore yield nearly the same self-energy. However, the density of states obtained within the FLF cluster scheme does not exhibit the pronounced coherence peaks at the gap edges that are characteristic of the mean-field density of states. Moreover, as the temperature increases, the spectral weight at the gap edges gradually increases. We interpret this behavior as the onset of pseudogap formation. Thus, Fig.~\ref{fig:lattice_DOS_near_Fermi} clearly shows the coexistence of the superconducting gap and the pseudogap over a finite temperature range. Upon further increasing the temperature, the gap in the FLF density of states closes and pseudogap begins to fill in. In other words, the spectral weight from the gap in the density of states is gradually redistributed toward the gap edges, because the sum rule for density of states is satisfied in FLF approach. As a result, at $T=0.116$ a well-defined pseudogap is already visible, which continues to fill in as the temperature  increase further.

A similar behavior of gap-filling and formation of a gap and a pseudogap was reported in Ref.~\cite{jiang2022monte} for a system of fermions coupled to $\text{SO}(2)$ quantum rotors. Also, this behavior is similar to the pseudogap formation in strongly-correlated systems with local correlations \cite{Peters2015local, Allen2001nonpertubative, Kyung2001pairing, Gauvin-Ndiaye2023improved, Martin2023Nonlocal}. In Ref.~\cite{stepanov2026superconductivity} the same two-gap picture was found for the $t–t'$ Hubbard model for a single correlated cuprate band, but short-range antiferromagnetic fluctuations are responsible in this case. However, our treatment is different because of the non-local nature of the FLF-approach. Coexisting  a gap and a pseudogap in the wide temperature range was seen in the strongly correlated regime where correlations are predominantly local. Our results show that such the phenomenon also arises in a moderate correlated regime with non-local correlations.

\FloatBarrier
\section{Conclusion}

In this work, we have demonstrated the formation of a pseudogap in the layered attractive Hubbard model using the fluctuating local field (FLF) method. This approach makes it possible to explicitly incorporate  superconducting fluctuations into the description of the system. In particular, it provides an effective description of the unstable collective mode with $\vec{K}=(0,0)$ and $\Omega=0$ for the $8\times8$ and $16\times16$ clusters. We have shown that this non-perturbative approach reproduces the mean-field results at low temperatures, remains free of divergences in the vicinity of the transition, and reproduces the results obtained from the summation of the zero-mode ladder diagrams at temperatures above $T_c$.

The cluster scheme further enables the description of the very large weakly coupled layers. Within this framework, a phase transition is observed as the temperature is increased. The density of states of the large lattice exhibits the coexistence of a superconducting gap and a pseudogap over a broad temperature range. Similar behaviour has previously been reported for fermions coupled to $\mathrm{SO}(2)$ quantum rotors~\cite{jiang2022monte}, as well as in the strongly correlated systems with predominantly local correlations~\cite{Peters2015local, Allen2001nonpertubative, Kyung2001pairing, Gauvin-Ndiaye2023improved, Martin2023Nonlocal}. At low temperatures, the density of states remains close to the mean-field result. As the temperature increases, however, a pseudogap develops and gradually fills in while the  gap closes. Thus, we have shown that the coexistence of a superconducting gap and a pseudogap can also occur in the moderately correlated systems with the predominantly non-local correlations.

\bibliography{Ref.bib}
\end{document}